# Electrons for Detection of Casimir Photons

Edik A. Ayryan[1], Karen G. Petrosyan[2],
Ashot H. Gevorgyan[3], Nikolay Sh. Izmailian[4], Koryun B. Oganesyan[1,4,*]

[1] Joint Institute for Nuclear Research, Dubna, Russia
[2] Institute of Physics, Academia Scinica, Taipei
[3] Yerevan State University, Yerevan, Armenia
[4] Alikhanyan National Science Lab, Yerevan Physics Institute, Yerevan, Armenia

[*] bsk@yerphi.am

**Abstract**
We propose a method for the detection of a dynamical Casimir effect. Assuming that the Casimir photons are being generated in an electromagnetic cavity with a vibrating wall (dynamical Casimir effect), we consider electrons passing through the cavity to be interacting with the intracavity field. We show that the dynamical Casimir effect can be observed via the measurement of the change in the average or in the variance of the electron's kinetic energy. We point out that the enhancement of the effect due to finite temperatures makes it easier to detect the Casimir photons.

## 1. INTRODUCTION

In 1948, Casimir predicted the existence of an attractive force between two perfectly conducting parallel plates placed in a vacuum [1]. Since then, a variety of fundamental and measurable consequences of quantum fluctuations under the influence of external conditions have been derived [2]. Recent precision measurements of the Casimir force [3] confirm the basic concepts of the quantum field theory in the presence of static external constraints. The success in measuring the static Casimir force intensified experimental efforts in order to verify a no less fundamental prediction, namely, the dynamical Casimir effect, i.e., the creation of particles out of a vacuum induced by the interaction with dynamical external constraints [4]. In particular, the creation of photons in vibrating cavities seems to be the most promising scenario for the possible experimental verification of motion-induced vacuum radiation [5]. Nowadays, the dynamical Casimir effect that appears due to the instability of the vacuum state of the electromagnetic field in the presence of time-dependent boundaries attracts increasing attention among theorists [6, 7] and experimentalists [8, 9]. The process involves the creation of Casimir photons in an electromagnetic cavity with a vibrating wall like electron vibration in free electron lasers [10-53]. This process can be effectively described by the following single-mode Hamiltonian [6]

$$H = \hbar a^+ a - i\hbar \frac{\lambda}{2}\left(a^2 e^{2i\omega t} - a^{+2} e^{-2i\omega t}\right), \qquad (1)$$

which shows a parametric generation of the photons. Here, $a^+$ and $a$ are the creation and annihilation operators for the photons at the frequency ω, respectively, and λ is the coefficient responsible for the parametric generation.
The detection of the Casimir photons with use of superradiance was recently proposed in [8]. The photons in the proposed setup are stored in a high-quality electromagnetic cavity and detected through their interaction with ultracold alkali-metal atoms prepared in an inverted population of hyperfine states. Superradiant amplification of the generated photons would result in a detectable radiofrequency signal temporally distinguishable from the expected background. An overview of this and related efforts can be found in [9]. In our paper, we propose another scheme for the detection of Casimir photons. We will exploit an interaction of the photons with *electrons* passing through the electromagnetic cavity with a vibrating wall. The interaction of an electron with an intracavity field while passing the cavity was actually considered half a century ago [54]. At that time, attention was paid to quantum fluctuations in the electron's kinetic energy and to relationships similar to Nyquist's formula. We will derive expressions for the average kinetic energy and its variance. However, our purpose is restricted to finding a way of detecting the photons generated in the dynamical Casimir effect.

As we will show, the dynamical Casimir effect can be observed via the measurement of the change in the average or in the variance of the electron's kinetic energy. Let us also point out that we present a fully quantum-mechanical approach. All of the variables for both the electron and the electromagnetic field are operators.

## 2. EQUATIONS OF MOTION

The 1D Hamiltonian for an electron interacting with an electromagnetic field is as follows [55]:

$$H = \frac{1}{2m}\left(p - \frac{e}{c}A(x,t)\right)^2, \qquad (2)$$

where $p$ is the momentum operator, $m$ is the mass of the electron, and $e$ is the electron's charge. $A(x,t)$ is the potential of the electromagnetic field. Let us consider the case when the potential of the electromagnetic field is only a function of time $A(x,t) = A(t)$ and does not depend on the position (this is the so-called dipole approximation), as is the case for the intracavity electron–photon interaction under consideration. In the dipole approximation, the Hamiltonian becomes equivalent to the following [56]:

$$H = \frac{p^2}{2m} - exE(t), \qquad (3)$$

where $x$ is the position operator and $E = -\frac{1}{c}\dot{A}(t)$ electric field. For the single-mode cavity, the electric field is

$$E(t) = \left(\frac{\hbar\omega}{2V\varepsilon_0}\right)^{1/2}\left(ae^{-i\omega t} + a^+e^{i\omega t}\right). \qquad (4)$$

Hence, we arrive at the following Hamiltonian that describes the interaction of the electron with the singlemode cavity quantized electromagnetic field:

$$H = \frac{p^2}{2m} - gx\left(ae^{-i\omega t} + a^+e^{i\omega t}\right), \qquad (5)$$

where $g = e\left(\dfrac{\hbar\omega}{2V\varepsilon_0}\right)^{1/2}$ is the coupling constant of the intracavity electron–photon interaction, and V is the volume of the cavity. In the interaction picture, where we replace $ae^{i\omega t} \to a$, one arrives at the following Hamiltonian, which describes the interaction of the electron passing the cavity with the intracavity electromagnetic field (the vibrating wall leads to the parametric generation of photons):

$$H = \frac{p^2}{2m} - gx\left(ae^{-2i\omega t} + a^+ e^{2i\omega t}\right) + \hbar a^+ a - i\hbar\frac{\lambda}{2}\left(a^2 - a^{+2}\right). \tag{6}$$

The equations of motion for the electron and the intracavity field are as follows:

$$\begin{aligned}
\dot{x} &= p/m, \\
\dot{p} &= g\left(ae^{-2i\omega t} + a^+ e^{2i\omega t}\right), \\
\dot{a} &= -i\omega a + \lambda a^+ + gxe^{2i\omega t}, \\
\dot{a}^+ &= i\omega a^+ + \lambda a + gxe^{-2i\omega t}.
\end{aligned} \tag{7}$$

We will drop the fast oscillating terms from the equations for the field operators $a$ and $a\dagger$ and will solve them, obtaining

$$\begin{aligned}
a(t) &= a(0)\cosh(\lambda t) + a^+(0)\sinh(\lambda t), \\
a^+(t) &= a(0)\sinh(\lambda t) + a^+(0)\cosh(\lambda t),
\end{aligned} \tag{8}$$

and

$$\begin{aligned}
N(t) &= a^+(t)a(t) = \sinh^2(\lambda t) + a^+ a\left(\cosh^2(\lambda t) + \sinh^2(\lambda t)\right) \\
&\quad + \left(a^2 + a^{+2}\right)\sinh(\lambda t)\cosh(\lambda t)
\end{aligned} \tag{9}$$

for the number of photons. Thus, we arrive at the problem of an electron moving in an external electromagnetic field without any influence of the electron on the intracavity field. Let us now make a comment related to the treatment of the problem of motion of an electron in an external fluctuating (thermal or quantum) electromagnetic field. As was shown by Klimontovich [57] (see also [58]), a Langevin equation approach to the problem may lead to incorrect results. A better way to treat the problem is to begin with the kinetic equation for the momentum probability distribution function.
This approach would allow us to treat systematically the process taking into account the conservation of both energy ($\hbar\omega = p^2(\tau)/2m - p^2(0)/2m$) and momentum ($\hbar k = p(\tau) - p(0)$). However, we restrict ourselves to the regime where we neglect the backaction of the electron on the field. We can now write down the equation for the momentum of the electron in the following form (hereafter, we will denote $a(0)$ and $a\dagger(0)$ as $a$ and $a\dagger$, respectively):

$$\frac{1}{8}\dot{p} = \left(a + a^+\right)e^{\lambda t}\cos\Omega t - i\left(a - a^+\right)e^{-\lambda t}\sin\Omega t, \tag{10}$$

with $\Omega = 2\omega$. The solution to this equation is as follows:

$$p(\tau) = p(0)$$
$$+ \frac{g}{\lambda^2 + \Omega^2}[(a + a^+)(\Omega e^{\lambda\tau} \sin\Omega t - \lambda(1 - e^{\lambda\tau}\cos\Omega t)) \tag{11}$$
$$-i(a - a^+)\Omega(1 - e^{-\lambda\tau}\cos\Omega t) - \lambda e^{-\lambda t}\sin\Omega t].$$

where $\tau$ is the flight time of the electron passing the cavity. We will assume that, initially, there was a thermal vacuum with temperature $T$. Hence, we have a zero value for the following quantities: $\langle a \rangle = \langle a^+ \rangle = \langle a^2 \rangle = \langle a^{+2} \rangle = 0$ and $\langle a^+ a \rangle = N_{th}$ the number of thermal photons, whereby $\langle ... \rangle$, the average over the initial state is assumed. For the average of the kinetic energy $K = \frac{p^2}{2m}$, we obtain

$$\langle K(\tau) \rangle = \langle K(0) \rangle + \frac{g^2(1 + 2N_{th})}{2m(\lambda^2 + \Omega^2)^2}(\Omega e^{\lambda\tau}\sin\Omega\tau - \lambda(1 - e^{\lambda\tau}\cos\Omega\tau))^2 \tag{12}$$
$$+ (\Omega(1 - e^{-\lambda\tau}\cos\Omega\tau) - \lambda e^{-\lambda\tau}\sin\Omega\tau)^2).$$

If the Casimir photons are absent, that is, in the case of a pure vacuum ($\lambda = 0$), we obtain

$$\langle K(\tau) \rangle = \langle K(0) \rangle + \frac{g^2(1 + 2N_{th})}{2m\omega^2}\sin^2\omega\tau \tag{13}$$

Notice that for the appropriately chosen parameters, such that $\omega\tau = \pi n$ ($n = 1, 2, ...$), the energy does not change. Meanwhile, for the same parameters and for $\lambda \neq 0$, we obtain the following change in the kinetic energy:

$$\Delta\langle K(\tau) \rangle = \frac{g^2(1 + 2N_{th})}{2m(\lambda^2 + \Omega^2)^2}[\lambda^2(e^{\lambda\tau} - 1)^2 + \Omega^2(1 - e^{-\lambda\tau})^2]. \tag{14}$$

having $\tau = 2\pi n/\Omega$, $n = 1, 2, ...$. For small values of $\lambda\tau \ll 1$, this simplifies to become

$$\Delta\langle K(\tau) \rangle = \frac{g^2(1 + 2N_{th})}{2m(\lambda^2 + (\Omega/2)^2)^2}. \tag{15}$$

Therefore, once one detects the change in the average kinetic energy for certain parameters, it would manifest the direct proof of the existence of Casimir photons. Let us now turn to a calculation of the kinetic energy variance $\langle(\Delta K(\tau))^2\rangle = \langle K^2(\tau) \rangle - \langle K(\tau) \rangle^2$., where we keep only terms on the order of $O(g^2)$

$$\langle(\Delta K(\tau))^2\rangle - \langle(\Delta K(\tau))^2\rangle = \frac{3g^2 \langle K(0) \rangle (1 + 2N_{th})}{2m(\lambda^2 + \Omega^2)^2} \tag{16}$$
$$\times((\Omega e^{\lambda\tau}\sin\Omega\tau - \lambda(1 - e^{\lambda\tau}\cos\Omega\tau))^2 + (\Omega(1 - e^{-\lambda\tau}\cos\Omega\tau) - \lambda e^{-\lambda\tau}\sin\Omega\tau)^2).$$

In the case of the absence of the Casimir photons ($\lambda = 0$), we get

$$\langle(\Delta K(\tau))^2\rangle = \langle(\Delta K(0))^2\rangle + \frac{g^2 \langle K(0)\rangle(1+2N_{th})}{2m(\lambda^2+(\Omega/2)^2)^2}\sin^2(\omega\tau/2). \tag{17}$$

Let us assume that the initial average kinetic energy is $\langle K(0)\rangle = mv_0^2/2$ and substitute it into the above equation

$$\langle(\Delta K(\tau))^2\rangle - \langle(\Delta K(0))^2\rangle + \frac{3}{4}g^2 L^2(1+2N_{th})\left(\frac{\sin^2(\omega\tau/2)}{\omega\tau/2}\right)^2. \tag{18}$$

where we introduced the length of the electron flight through the cavity $L = v_0\tau$. From this equation, we notice once again that, for the parameters such that $\Omega\tau = 2\pi n$, the variance of the kinetic energy remains constant. Meanwhile, for the same parameters and for small enough $\lambda\tau \ll 1$, we obtain

$$\langle(\Delta K(\tau))^2\rangle - \langle(\Delta K(0))^2\rangle = \frac{3}{4}\frac{g^2 L^2(1+2N_{th})}{1+\Omega^2/\lambda^2}. \tag{19}$$

Now, we have two ways of checking for the presence of Casimir photons. In one way, we may employ the change of the average kinetic energy and, in the other, we should exploit the kinetic energy variance.

### 3. CONCLUSION

In conclusion, we have proposed an experimental way for detecting a dynamical Casimir effect. The setup involves the interaction of electrons passing a cavity with a vibrating wall and interacting with the intracavity electromagnetic field. For certain parameters, it becomes possible to directly recognize the presence of Casimir photons. We point out that the enhancement of the effect for finite temperatures makes it easier to detect the dynamical Casimir effect.